\documentclass[
aps,
prb,
showpacs,
floatfix,
reprint,
twoside,
superscriptaddress
]{revtex4-1}
\usepackage{amsmath}
\usepackage[T1]{fontenc}
\usepackage{fourier}
\usepackage{graphicx}
\usepackage[colorlinks=true,
allcolors=blue,
dvipdfm=true]{hyperref}

\begin{document}

\title{Chirality and spin transformation of triplet Cooper pairs upon interaction with singlet condensate}

\author{Andreas Moor}
\affiliation{Theoretische Physik III, Ruhr-Universit\"{a}t Bochum, D-44780 Bochum, Germany}
\author{Anatoly F.~Volkov}
\affiliation{Theoretische Physik III, Ruhr-Universit\"{a}t Bochum, D-44780 Bochum, Germany}
\author{Konstantin B.~Efetov}
\affiliation{Theoretische Physik III, Ruhr-Universit\"{a}t Bochum, D-44780 Bochum, Germany}
\affiliation{National University of Science and Technology ``MISiS'', Moscow, 119049, Russia}

\begin{abstract}
We show that the fully polarized triplet s\nobreakdash-wave component is characterized not only by the spin direction, but also by chirality. Interaction of a polarized triplet component and a singlet one results in creation of triplet Cooper pairs with opposite spin direction or of different chiralities. Such spin transformation leads to interesting phenomena in multiterminal magnetic Josephson junctions. We calculate the dc Josephson current~$I_{\text{J}}$ in a multiterminal Josephson contact of the S$_{\text{m}}$/n/S$_{\text{m}}^{\prime}$ type with ``magnetic'' superconductors~S$_{\text{m}}$ that generate fully polarized triplet components. The superconductors~S$_{\text{m}}$ are attached to magnetic insulators (filters) which let to pass electrons with a fixed spin direction only. The filter axes are assumed to be oriented antiparallel to each other. The Josephson current is zero in two-terminal Josephson junction, i.e., in~S/n/S$_{\text{m}}$ or in~S$_{\text{m}}$/n/S$_{\text{m}}^{\prime}$ contact. But in the three-terminal Josephson junction, with another S~superconductor attached to the normal wire, the finite current~$I_{\text{J}}$ appears flowing from the S~superconductor to S$_{\text{m}}$~superconductors. The currents through the right (left) superconductors~S$_{\text{m}}$ are opposite in sign, ${I_{\text{R}} \equiv I_{\text{J}} = I_{\text{c}} \sin ( \chi_{\text{R}} + \chi_{\text{L}} - 2 \chi ) = - I_{\text{L}}}$, where~$\chi_{\text{L/R}}$ and~$\chi$ are the phases of superconductors~S$_{\text{m}}$,~S$_{\text{m}}^{\prime}$, and~S, respectively. We discuss possibilities of experimental observation of the effect.
\end{abstract}

\date{\today}

\maketitle

\section{Introduction.}

Most of superconducting materials are singlet superconductors, i.e., Cooper pairs in these materials consists of two electrons with opposite spins and momenta. The wave function of the Cooper pairs~$f$ and the superconducting order parameter~$\Delta$ has the same sign for any direction of mometum~$\mathbf{p}$ ($s$\nobreakdash-type superconductivity) in the low~T$_{\text{c}}$ superconductors or change sign by rotation of the~$\mathbf{p}$ vector by~$\pi/2$ ($d$\nobreakdash-type superconductivity) in high~T$_{\text{c}}$ superconductors.\cite{HTSrev95,HTSrev00}

Another type of pairing---the triplet one---has been observed first in the quantum liguid $^{3}$He.\cite{vollhardt,volovik,Mineev} As to superconductors, only in the last decades it has been established that the triplet superconductivity may arise in solids. Namely, it is believed that the $p$\nobreakdash-wave triplet Cooper pairs may exist in heavy-Fermion compounds (UPt$_{3}$), in quasi-one dimensional organic materials and in crystals of Sr$_{2}$RuO$_{4}$ (see the Reviews Refs.~\onlinecite{Sigrist_Ueda_1991,Mackenzie_RevModPhys.75.657,Mineev}). In these superconductors, the wave function~$f(\mathbf{p})$ and the order parameter~$\Delta(\mathbf{p})$ are odd functions of the momentum~$\mathbf{p}$. Hence, the scattering on conventional impurities and on interfaces or rough surfaces leads to isotropisation of these functions and finally to the suppression of the superconductivity. Thus, the triplet $p$\nobreakdash-wave superconductivity takes place only in sufficiently clean samples. Perhaps this is the reason why numerous theoretical predictions concerning the Josephson and proximity effects in the systems TS-SS, TS-TS and TS-F-S,\cite{Kastening_et_al_2006,Asano_et_al_2007_a,Sengupta_Yakovenko_2008,Lu_Yip_2009,Linder_Cuoco_Sudbo_2010,Terrade_Gentile_Cuoco_Manske_2013,Brydon_Chen_Asano_Manske_2013} have yet not been detected in experiments which usually deal with diffusive materials (here, TS stands for a $p$\nobreakdash-wave triplet superconductor, while SS represents a singlet superconductor, and~F---a ferromagnet).

On the other hand, isotropic triplet superconductivity (or, to be more exact, triplet superconducting correlations) may arise in a conventional singlet superconductor in the presence of an external magnetic field~$\mathbf{H}$ or internal exchange field~$\mathbf{h}$ (see reviews Refs.~\onlinecite{BuzdinRMP,BVErmp,Eschrig_Ph_Today} and references therein). In a homogeneous~$\mathbf{H}$ or~$\mathbf{h}$, the total spin of triplet Cooper pairs~$\mathbf{S}$ is oriented perpendicularly to the vector~$\mathbf{H}$, respectively~$\mathbf{h}$, so that spins of the condensate do not contribute to the induced magnetic moment (zero Knight shift)~\cite{Knight,Abrikosov_Gorkov,Larkin_Ovchinnikov_1965,Fulde_Ferrell_1964}. However, if the field~$\mathbf{h}(x)$ is nonuniform, the triplet component with ${\mathbf{S} \parallel \mathbf{h}}$ may arise in the system. Such a case is realized, for instance, in multilayered F$_{1}$/F$_{2}$/S structures with noncollinear exchange fields~$\mathbf{h}_{1,2}$ in the ferromagnets~F$_{1,2}$. The singlet component penetrates into the ferromagnetic layers F$_{1,2}$ provided that~F$_{2}$ is a weak ferromagnet or the F$_{2}$~film is sufficiently thin. Under the action of the exchange field~$\mathbf{h}_{2}$, the singlet component creates triplet Cooper pairs with zero projection of the vector~$\mathbf{S}_{0}$ on~$\mathbf{h}_{2}$. If the vectors~$\mathbf{h}_{1,2}$ are not collinear, some triplet Cooper pairs have a projection of the vector~$\mathbf{S}_{\pm 1}$ parallel to the direction of~$\mathbf{h}_{1}$. The singlet Cooper pairs and the triplet pairs with zero projection of~$\mathbf{S}$ on~$\mathbf{h}_{1}$ penetrate into the diffusive~F$_{1}$ over a short distance ${\xi_{h} \propto 1 / \sqrt{h_{1}}}$, whereas the penetration depth of the Cooper pairs with ${\mathbf{S} \parallel \mathbf{h}_{1}}$ is rather long and may reach the value ${\xi_{T} \propto 1 / \sqrt{T}}$ (usually ${T \ll h}$)~\cite{Bergeret_Volkov_Efetov_2001}.

In the first case, the exchange field~$\mathbf{h}_{1}$ destroys the singlet Cooper pairs, but does not affect the triplet Cooper pairs with total spin~$\mathbf{S}$ parallel to the vector~$\mathbf{h}_{1}$. The component of the wave function describing the latter Cooper pairs is called the long-range triplet component (LRTC)~\cite{BVErmp,Eschrig_Ph_Today}.

Very important for the experimental observation of the LRTC is the independence of the penetration length~$\xi_{T}$ on the exchange field~$\mathbf{h}$ and a weak dependence of this length on the impurity concentration. To be more exact, this dependence is nearly the same as the corresponding dependence of penetration length of the singlet component into a normal metal, i.e., in the ballistic case this length is~${\xi_{h} \simeq v_{\text{F}}/T}$ and in the diffusive case---${\xi_{h} \simeq \sqrt{D/T}}$, where ${D = v_{\text{F}} l / 3}$ is the diffusion coefficient with the Fermi velocity~$v_{\text{F}}$. These circumstances allow one to detect the LRTC easily and to manipulate it.

Indeed, the effect of the deep penetration of the LRTC into strong ferromagnets has been observed in several experiments. In most of these experiments, the Josephson current~$I_{\text{J}}$ has been measured in S/F JJs of different types. Some JJs were S/F$_{1}$\ldots{}F$_{n}$/S multilayered structures with noncollinear magnetizations~$\mathbf{M}_{i}$ in different F$_{i}$~layers~\cite{Birge10,*Birge12,Zabel10}. In the case of collinear magnetizations~$\mathbf{M}_{i}$, the current has been observer as being negligible and increased drastically if the LRTC appeared due to noncollinearity of~$\mathbf{M}_{i}$~\cite{Birge10,*Birge12,Zabel10}. In other experiments, single layer S/F/S JJs were used with the magnetization vector~$\mathbf{M}$ in~F that changed its direction in space (helical ferromagnet)~\cite{Petrashov06,Petrashov11,Blamire10,*Blamire12}. In the third group of experiments~\cite{Keizer06,Aarts10,*Aarts11,*Aarts12}, the LRTC has been created eventually at the so-called spin-active interfaces. The LRTC resulted in a measurable Josephson current. In experiments by Leksin~\emph{et al.}\cite{Leksin_et_al_2012} the dependence of the critical temperature~T$_{\text{c}}$ on the angle between the magnetization vectors~$\mathbf{M}$ and~$\mathbf{M}^\prime$ in ferromagnets~F and F$^\prime$ in the system F/F$^\prime$/S has been observed. This dependence is a result of the appearance of the LRTC.

All types of JJs where the LRTC arises were investigated theoretically in detail, i.e., the current~$I_{\text{J}}$ has been calculated for multilayered S/F$_{1}$\ldots{}F$_{n}$/S JJs with spin-inactive interfaces~\cite{VE10,Buzdin07,Radovic10,Meng13,Buzdin15,Fominov07,Braude07} and for S/F/S JJs with spin-active interfaces~\cite{Eschrig03,Eschrig05,Eschrig07,Golubov_et_al_2009,Tanaka07b,Eschrig08,Zaikin08,Valls08,Valls12,Eschrig13,Beenakker09,Brouwer11,Linder09,Linder10}.

The proximity effect in S/F structures\cite{Bergeret_Volkov_Efetov_2001,Kad01,Asano_et_al_2007,Linder_et_al_2009,Beenakker09,Linder09,Zhu_et_al_2010,Knezevic_et_al_2012} with nonhomogeneous magnetization and the critical temperature of the superconducting transition in multilayered S/F/F$^\prime$ structures\cite{Fominov_Golubov_Kupriyanov_JETP_Lett_2003,Fominov_et_al_JETP_Lett_2010} have also been studied theoretically.

It is important that in both cases the LRTC are, generally speaking, different. One can consider spin-active interface as a superconducting film~S covered by a thin layer of a weak ferromagnet~F$_{\text{w}}$ and attached to a thin magnetic insulator~I$_{\text{m}}$; we denote this system as~S$_{\text{m}}$/I$_{\text{m}}$ (corresponding to the structure S/F$_{\text{w}}$/I$_{\text{m}}$). If electrons with only one spin direction (say, oriented in the direction of the $z$\nobreakdash-axis) can tunnel through the magnetic insulator and the exchange field in~F$_{\text{w}}$ is not parallel to the $z$\nobreakdash-axis, then the triplet Cooper pairs penetrating into the n~or F~wire in an S$_{\text{m}}$/n (respectively, S$_{\text{m}}$/F) structure would be fully polarized, i.e., they would have only~$S_{\uparrow \uparrow}$ component of the total spin. On the other hand, the triplet Cooper pairs penetrating into the n~or F~wire through a strong ferromagnet~F$_{\text{s}}$ in an S/F$_{\text{w}}$/F$_{\text{s}}$/n or S/F$_{\text{w}}$/F$_{\text{s}}$/F structure have both projections,~$S_{\uparrow \uparrow}$ and $S_{\downarrow \downarrow}$, of the total spin. The presence of differently oriented Cooper pairs in the S/F$_{\text{w}}$/F$_{\text{s}}$/n or S/F$_{\text{w}}$/F$_{\text{s}}$/F structure is caused by the fact that the Zeeman term in the Hamiltonian describing the interaction of the condensate and the exchange field field in the strong ferromagnet, F$_{\text{s}}$, commutes with the wave (Green's) function of these Cooper pairs.

Different types of the triplet Cooper pairs manifest themselves in JJs of two types, S$_{\text{m}}$/I$_{\text{m}}$/n/S$_{\text{m}}$/I$_{\text{m}}$/S$_{\text{m}}$ and S$_{\text{m}}$/F$_{\text{s}}$/n/F$_{\text{s}}$/S$_{\text{m}}$, leading to quite different behavior of the Josephson current~$I_{\text{J}}$ as a function of relative polarization of filters or exchange field~$\mathbf{h}_{\text{s}}$ in~F$_{\text{s}}$ at the right and left banks of the JJ~\cite{Moor_2015_arxiv}. Whereas the current~$I_{\text{J}}$ has the same value for parallel or antiparallel orientations of the vectors of the exchange field~$\mathbf{h}_{\text{s}}$ in the right or left ferromagnet~F$_{\text{s}}$ in S$_{\text{m}}$/F$_{\text{s}}$/n/F$_{\text{s}}$/S$_{\text{m}}$ JJ, the current~$I_{\text{J}}$ differs from zero at equally oriented filter axes~I$_{\text{m}}$ in S$_{\text{m}}$/I$_{\text{m}}$/n/I$_{\text{m}}$/S$_{\text{m}}$ JJs and equals zero, ${I_{\text{J}} = 0}$, at antiparallel orientations of the filter axes~\cite{Moor_2015_arxiv}. This means that the quantum interference occurs only between parallel oriented triplet Cooper pairs. No interference takes place for the triplet Cooper pairs with~$S_{\uparrow \uparrow}$ and $S_{\downarrow \downarrow}$ total spins. Note that as a filter one can use not only magnetic insulator, but also a magnetic half metal. The latter also lets to pass only fully polarized Cooper pairs. The penetration of the triplet component through a half metal has been studied in Refs.~\onlinecite{Eschrig03,Tanaka07b,Buzdin15_arXiv}.

It is important to note that in the case of antiparallel filter axes not only Cooper pairs can not flow through the filters, but also the quasiparticle current is zero. Therefore, the S$_{\text{m}}$/I$_{\text{m} \downarrow}$/n/I$_{\text{m} \uparrow}$/S$_{\text{m}}$~system at any temperatures is not penetrable for charge carriers of any type (arrows denote the direction of filter axes).

In this Paper, we study the interaction between fully polarized triplet Cooper pairs with the singlet condensate in a Josephson junctions of the S$_{\text{m}}$/I$_{\text{m} \downarrow}$/n/I$_{\text{m} \uparrow}$/S$_{\text{m}}$-type to which a conventional singlet BCS superconductor is attached via a normal n~wire [see Fig.~\ref{fig:setup}~(a)].

In the assumed case of antiparallel filter axes, the  Josephson current is zero as is the case in an S/n/I$_{\text{m}}$/S$_{\text{m}}$-type contact in the same order of approximation in the transmission coefficient of the n/I$_{\text{m}}$/S$_{\text{m}}$~interface. We show that the interaction of the singlet and triplet components in an S$_{\text{m}}$/I$_{\text{m} \downarrow}$/n/S/n/I$_{\text{m} \uparrow}$/S$_{\text{m}}$-type Josephson contact with antiparallel filter axes leads to a nonzero Josephson current, ${I_{\text{J}} \neq 0}$.

We calculate the Josephson current~$I_{\text{J}}$ in the lowest order of approximation in the transmittance coefficient~$r_{\text{R},\text{L}}$ of the right~(left) n/I$_{\text{m}}$/S$_{\text{m}}$ interfaces and show that in the three-terminal S$_{\text{m}}$/I$_{\text{m} \downarrow}$/n/S/n/I$_{\text{m} \uparrow}$/S$_{\text{m}}$~JJ with antiparallel I$_{\text{m} \uparrow,\downarrow}$~filters, the Josephson current~$I_{\text{J}}$~(${I_{\text{J}} \propto r_{\text{L}} r_{\text{R}}}$) is finite only if all three terminals are present. On the other hand,~${I_{\text{J}} = 0}$ in any two-terminal JJs, i.e., in S$_{\text{m}}$/I$_{\text{m} \downarrow}$/n/I$_{\text{m} \uparrow}$/S$_{\text{m}}$ and in S/n/I$_{\text{m}}$/S$_{\text{m}}$ junctions.

Moreover, we show that the triplet components with~$\pm 1$ projection of the total spin~$\mathbf{S}$ on the $z$~axis are characterized not only by the spin projection,~$+1$ or $-1$, but also by different chiralities. This means that spin-up (spin-down) triplet Cooper pairs are described by two independent wave functions of different chiralities, which has physical consequences manifesting itself, e.g., in different expressions for the Josephson current.

\section{System and main equations.}

\begin{figure}
  \centering
  \includegraphics[width=1.0\columnwidth]{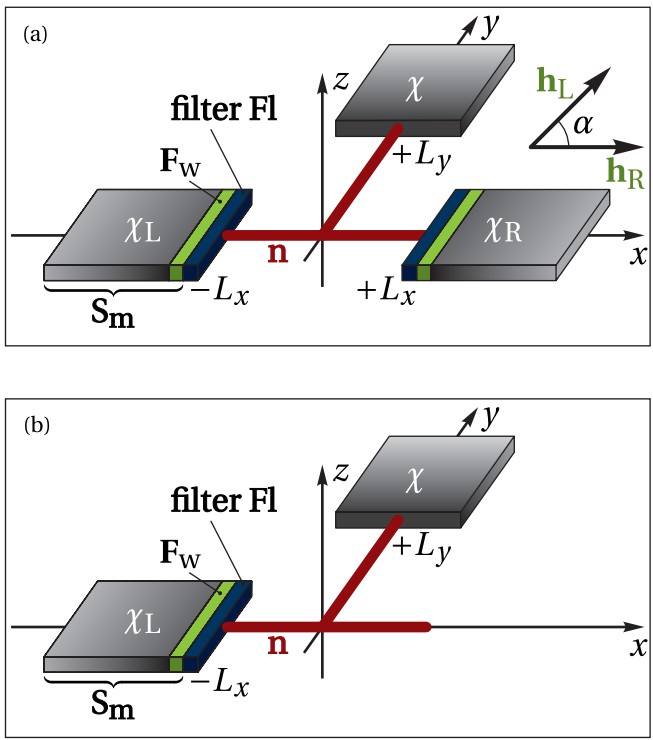}
  \caption{Schematic representation of the systems under consideration. (a)~The TST~contact. (b)~The TS~contact.} \label{fig:setup}
\end{figure}

We consider two types of ``magnetic'' Josephson junctions, which we denote as ``TST''~(triplet-singlet-triplet) and, respectively, ``TS''~(triplet-singlet) contacts [see corresponding Figs.~\ref{fig:setup}~(a) and~\ref{fig:setup}~(b)]. The TST contact is a S$_{\text{m}}$/Fl/n/S/n/Fl$^{\prime}$/S$_{\text{m}}^{\prime}$ junction with a tunnel spin-active barrier~Fl~(filter) with a spin-dependent transparency (such kinds of magnetic insulators have been already used in experiments, e.g., ultrathin EuO~films~\cite{Moodera08} or DyN and GdN~\cite{Muduli_et_al_arXiv_2015}) separating the n~wire from two~S$_{\text{m}}$ and~S$_{\text{m}}^{\prime}$ reservoirs that represent superconductors with exchange fields~$\mathbf{h}_{\text{R},\text{L}}$ lying in the $(x,y)$~plane.

The ``magnetic'' superconductor~S$_{\text{m}}$ may be realized in the form of a superconductor/weak ferromagnet bilayer, S/F$_{\text{w}}$, with an effective exchange field ${\mathbf{h} = \mathbf{h}_{\text{F}} d_{\text{F}} / (d_{\text{F}} + d_{\text{S}})}$, where~$\mathbf{h}_{\text{F}}$ is an exchange field in~F$_{\text{w}}$ and~$d_{\text{F},\text{S}}$ are the thicknesses of the~F$_{\text{w}}$ and S~films, respectively~\cite{Bergeret_Volkov_Efetov_2001_b,Moor_2015_arxiv}. One more superconductor, which is a conventional singlet superconductor~S, is attached to the normal wire (or film). Penetration of Cooper pairs through~Fl is taken into account via boundary conditions. We consider the case when the filter~Fl passes electrons with only one spin direction collinear with the $z$~axis. The TS contact represents a TST contact with detached~S$_{\text{m}}^{\prime}$ reservoir.

As in Refs.~\onlinecite{BVErmp,Moor_2015_arxiv}, we employ the well developed method of quasiclassical Green's functions~\cite{RammerSmith,LO,BelzigRev,Kopnin}. In particular, this technique has been used succesfully in describing superconducting mesoscopic heterostructures.\cite{Volkov199321,ZAITSEV1994274,Nazarov_1994,Hekking_Hekking_1993,Hekking_Hekking_1994,Nazarov19991221} In the considered case of singlet and triplet Cooper pairs, the quasiclassical Green's functions~$\hat{g}$ are matrices in the particle-hole and spin spaces with basis represented by tensor products ${\hat{X}_{ik} = \hat{\tau}_{i} \cdot \hat{\sigma}_{k}}$, where the Pauli matrices~$\hat{\tau}_{i}$ and~$\hat{\sigma}_{k}$ (${i,k=1,2,3}$) and the unit matrices~$\hat{\tau}_{0}$ and~$\hat{\sigma}_{0}$ operate in the particle-hole and spin space, respectively (see also Ref.~\onlinecite{BVErmp}).

We consider the most relevant from experimental point of view diffusive structures, where the mean free path~$l$ is much shorter than the coherence length~${\xi_{T} = \sqrt{D/\pi T}}$ with the diffusion coefficient ${D = v_{\text{F}} l/3}$, the Fermi velocity~$v_{\text{F}}$ and the temperature~$T$. Only s-wave Cooper pairs, triplet or singlet, survive in this case.

In the n~wires in the $\nu$\nobreakdash-direction (${\nu = x,y}$), these matrix Green's functions~$\hat{g}(\omega)$ in the Matsubara representation obey the Usadel equation~\cite{Usadel,BVErmp,Moor_2015_arxiv}
\begin{equation}
   - \partial_{\nu}(\hat{g} \partial_{\nu} \hat{g}) + \kappa_{\omega}^{2} [\hat{X}_{30} \,, \hat{g}] = 0 \,,
    \label{1}
\end{equation}
and the normalization condition
\begin{equation}
\hat{g} \cdot \hat{g} = 1 \,, \label{eq:norm_cond}
\end{equation}
where ${\kappa_{\omega}^2 = |\omega| / D}$, and ${\omega = (2n+1) \pi T}$ is the Matsubara frequency.~\footnote{Note that the representation of the matrix Green's functions~$\hat{g}$ differs somewhat from those ~$\hat{g}_{\text{BVE}}$ used in Refs.~\onlinecite{Bergeret_Volkov_Efetov_2001,BVErmp}. We employed the transformation suggested by Ivanov and Fominov~\cite{IvanovFomin} so that ${\hat{g} = \hat{U} \cdot \hat{g} \cdot \hat{U}^{\dagger}}$ with ${\hat{U} = (1/2)(1 + i \hat{X}_{33}) \cdot (1 - i \hat{X}_{03})}$}

At~$\pm L_{\nu}$ the boundary conditions can be written as (see Refs.~ \onlinecite{EschrigBC13,*EschrigBC13a,EschrigBC15} and Eq.~(4.7) in Ref.~\onlinecite{Bergeret12b})
\begin{equation}
    L_{\nu} \hat{g} \partial_{\nu} \hat{g}_{|_{\nu = \pm L_{\nu}}} = \pm r_{\nu} [\hat{g} \,, \hat{\mathrm{\Gamma}}_{\nu} \hat{G}_{\nu} \hat{\mathrm{\Gamma}}_{\nu}]_{|_{\nu = \pm L_{\nu}}} \,,
    \label{2}
\end{equation}
where ${r_{\nu} = L_{\nu}/\sigma R_{\text{b},\nu}}$, with the conductivity of the n~wire~$\sigma$ and the n\nobreakdash-Fl interface resistance at~$\pm L_{\nu}$ per unit area~$R_{\text{b},\nu}$ assumed to be equal for left and right banks.

At the interface between the normal wire and the singlet superconductor, the matrix~$\hat{\mathrm{\Gamma}}_{\nu}$ does not depend on spins, such that ${\hat{\mathrm{\Gamma}}_{\nu} = 1}$. In contrast, at the S$_{\text{m}}$/Fl/n~interface, the matrix transmission coefficient~$\hat{\mathrm{\Gamma}}_{\nu}$ describes the electron transmission with a spin-dependent probability~$\mathcal{T}_{\uparrow, \downarrow}$. If the filters let to pass only electrons with spins aligned parallel to the $z$~axis, then ${\hat{\mathrm{\Gamma}}_{\nu} = \mathcal{T}_{\nu} + \mathcal{U}_{\nu} \hat{X}_{33}}$ so that the probability for an electron with spin up (down) to pass into the n~wire is ${\mathcal{T}_{\uparrow, \downarrow} = \mathcal{T}_{\nu} + \zeta_{\nu} \mathcal{U}_{\nu}}$ with  ${\zeta_{\nu} = +1}$ if the filter passes only spin-up electrons, whereas ${\zeta_{\nu} = -1}$ means that the filter passes spin-down electrons only. Therefore, the condition ${\mathcal{T}_{\nu} = \zeta_{\nu} \mathcal{U}_{\nu}}$ means that electrons with only one spin orientation are allowed to pass through the filter. The coefficients~$\mathcal{T}$ and~$\mathcal{U}$ are assumed to be normalized, i.e., ${\mathcal{T}^{2} + \mathcal{U}^{2} = 1}$.

Quasiclassical Green's function matrices~$\hat{G}_{\nu}$ in superconductors~S$_{\nu}$ in presence of a uniform exchange field~$\mathbf{h}_{\nu}$ oriented along the $z$~axis [${\mathbf{h}_{\nu} = (0, 0 , h_{\nu})}$] have the form (dropping~$\nu$ for brevity)
\begin{equation}
    \hat{G}_0 = g_{+} \hat{X}_{30} + g_{-} \hat{X}_{33} + \hat{F} \,. \label{3}
\end{equation}
Here, the first two terms describe the normal part, and the last term,
\begin{equation}
\hat{F} = \hat{X}_{10} f_{+} + \hat{X}_{13} f_{-} \,, \label{F}
\end{equation}
is the anomalous (Gor'kov's) part, where ${f_{\pm} = (1/2) [f(\omega + i h) \pm f(\omega - i h)]}$ with ${f(\omega) = \Delta / \sqrt{\omega^{2} + \Delta^{2}}}$. The functions~$g_{\pm}$ are obtained from~$f_{\pm}$ using the orthogonality condition Eq.~(\ref{eq:norm_cond}).

The first, respectively, the second terms in the expression for~$\hat{F}$, Eq.~(\ref{F}) correspond to the singlet component, respectively, to triplet component with zero projection of the total spin of a Cooper pair on the direction of the vector~$\mathbf{h}$. The triplet component~$f_{-}$ turns to zero at ${h = 0}$ (in the singlet superconductor) and, as should be in accordance with the Pauli principle, is an odd function of~$\omega$.

Presence of superconducting phase~$\chi$ is introduced via a gauge transformation ${\hat{F}_{\chi} = \hat{S}_{\chi} \cdot \hat{F} \cdot \hat{S}_{\chi}^{\dagger}}$, with the unitary matrix ${\hat{S}_{\chi} = \exp[ \hat{X}_{30} i \chi / 2]}$. Moreover, with the help of the rotation matrix
\begin{equation}
\hat{R}_{\beta, j} = \cos(\beta/ 2) + i \sin(\beta/ 2) \hat{X}_{0j} \label{eq:rotation_matrix}
\end{equation}
it is easy to obtain the matrix~$\hat{F}_{\beta}$ for an arbitrary orientation of the exchange field ${\mathbf{h} = \{ h_{i} \}}$.

In the following, we concentrate on the case when the exchange fields~$\mathbf{h}_{\text{R},\text{L}}$ in the tight and left superconductors are perpendicular to the magnetization~$\mathbf{M}_{\text{s}}$ in the strong ferromagnet, i.e.,~${\mathbf{h}_{\text{R},\text{L}} \perp \mathbf{M}_{\text{s}}}$. In this case, the triplet components at the right and left have no nonzero projection on the filter axes (parallel, respectively, antiparallel with the $z$~axis). Thus, the Green's function in the $\nu$\nobreakdash-th superconductor with the phase~$\chi_{\nu}$ and with the exchange field lying in the $(x,y)$~plane and setting up the angle~$\alpha_{\nu}$ with the $x$~axis reads
\begin{equation}
\hat{G}_{\nu} = \hat{R}_{\alpha_{\nu}, 3} \hat{R}_{-\pi/2, 2} \hat{S}_{\chi_{\nu}} \hat{G}_0 \hat{S}_{\chi_{\nu}}^{\dagger} \hat{R}_{-\pi/2, 2}^{\dagger} \hat{R}_{\alpha_{\nu}, 3}^{\dagger} \,, \label{G_nu}
\end{equation}
and we choose the Green's functions in the right superconductor as
\begin{equation}
\hat{G}_{\text{R}} \equiv \hat{G}_{\text{R}}(\chi_{\text{R}}) = \hat{R}_{-\pi/2, 2} \hat{S}_{\chi_{\text{R}}} \hat{G}_0 \hat{S}_{\chi_{\text{R}}}^{\dagger} \hat{R}_{-\pi/2, 2}^{\dagger} \,, \label{eq:G_R}
\end{equation}
i.e., we set the angle~${\alpha = 0}$ in the right superconductor and the Green's function in the left superconductor is obtained by exchange ${\chi_{\text{R}} \to \chi_{\text{L}}}$ and a rotation of~$\hat{G}_{\text{R}}$ around the $z$~axis by the angle~$\alpha$,
\begin{equation}
\hat{G}_{\text{L}} \equiv \hat{G}_{\text{L}}(\chi_{\text{L}}) = \hat{R}_{\alpha, 3} \hat{R}_{-\pi/2, 2} \hat{S}_{\chi_{\text{L}}} \hat{G}_0 \hat{S}_{\chi_{\text{L}}}^{\dagger} \hat{R}_{-\pi/2, 2}^{\dagger} \hat{R}_{\alpha, 3}^{\dagger} \,. \label{eq:G_L}
\end{equation}

Solving the Eq.~(\ref{1}) supplemented with Eqs.~(\ref{eq:norm_cond}) and~(\ref{2}), and using expressions in Eqs.~(\ref{eq:G_R}) and~(\ref{eq:G_L}), we can calculate the Josephson~$I_{\text{Q}}$ and spin~$I_{\text{sp}}$ currents through the interface~${\nu = \pm L_{\nu}}$ which are given by the expressions
\begin{align}
    I_{\text{Q},\nu} &= \sigma e^{-1} 2 \pi i T \sum_{\omega} \mathrm{Tr}\{ \hat{X}_{30} \cdot \hat{g} \partial_{x} \hat{g} \}_{|{\nu}}
    \label{4} \,, \\
    I_{\text{sp},\nu} &= \mu_{\text{B}} \sigma e^{-2} 2 \pi i T \sum_{\omega} \mathrm{Tr}\{ \hat{X}_{03} \cdot \hat{g} \partial_{x} \hat{g} \}_{|{\nu}} \,,
    \label{4a}
\end{align}
where $\mu_{\text{B}}$~is the effective Bohr magneton and $e$\nobreakdash---the elementary charge. Moreover, we can determine the magnetic moment in the n~wire ${\mathbf{M} = \{ M_i \}}$,
\begin{equation}
M_i = \mu_{\text{B}} \mathcal{N} 2 \pi T \sum_{\omega} \mathrm{Tr}\{ \hat{X}_{3i} \cdot \hat{g} \} \,, \label{eq:magn_moment}
\end{equation}
where~$\mathcal{N}$ is the density of states in the normal state.

The quasiclassical Green's function ${\hat{g} = \hat{g}_{\text{n}} + \hat{f}}$ consists of the normal Green's function~$\hat{g}_{\text{n}}$ which is diagonal in the particle-hole space, and of the nondiagonal condensate function~$\hat{f}$. The solution can easily be found for the case of short n~wires. We will be interested only in the case ${\mathcal{T} = \pm \mathcal{U}}$ when the singlet component does not penetrate through the filters.

Assuming short n~wires (${L_{\nu} \ll \xi_{T} \equiv \sqrt{D/\pi T}}$) and integrating Eq.~(\ref{1}) over~$\pm L_{\nu}$ taking into account the boundary conditions Eq.~(\ref{2}), we arrive at the equation~\cite{Mai13}
\begin{equation}
    [\hat{\mathrm{\Lambda}} \,, \hat{g}] = 0 \,, \label{5}
\end{equation}%
where ${\hat{\mathrm{\Lambda}} = \hat{\mathrm{\Lambda}}_{0} + \delta \hat{\mathrm{\Lambda}}}$, with ${\hat{\mathrm{\Lambda}}_{0} = (\kappa_{\omega} L)^{2} \hat{X}_{30} + \hat{\mathrm{\Lambda}}_{y}}$, ${L = 2 L_x + L_y}$, ${\delta \hat{\mathrm{\Lambda}} = \hat{\mathrm{\Lambda}}_{x}}$, and
\begin{equation}
\hat{\mathrm{\Lambda}}_{\nu} = [r_{\nu} \hat{\mathrm{\Gamma}}_{\nu} \hat{G}_{\nu} \hat{\mathrm{\Gamma}}_{\nu} ]_{|_{+L_{\nu}}} + [r_{\nu} \hat{\mathrm{\Gamma}}_{\nu} \hat{G}_{\nu} \hat{\mathrm{\Gamma}}_{\nu} ]_{|_{-L_{\nu}}} \,. \label{eq:Lambda_nu}
\end{equation}
In Eq.~(\ref{eq:Lambda_nu}), for~$\hat{\mathrm{\Lambda}}_{y}$ the second term vanishes, and for~$\hat{\mathrm{\Lambda}}_{x}$ in the TS~contact---the first term.

We assume that the interface resistances~$R_{\text{b},x}$ are larger than the resistances~$L_{x} / \sigma_{x}$ so that ${r_{x} \ll 1}$ and we can easily find solutions of Eq.~(\ref{5}) using an expansion~${\hat{g} = \hat{g}_{0} + \hat{g}_{1} + \ldots}$, where the zeroth order~$\hat{g}_{0}$ is obtained from Eq.~(\ref{5}) employing the normalization condition ${\hat{g} \cdot \hat{g} = 1}$,
\begin{equation}
\hat{g}_{0} = \mathcal{E}^{-1} \hat{\mathrm{\Lambda}}_{0} = \mathcal{E}^{-1} \big[ \tilde{G}_{\text{S}} \hat{X}_{30} + \hat{F}_{\text{S}} \big] \,, \label{g_0}
\end{equation}
with ${\mathcal{E}^{2} = \mathrm{Tr}(\hat{\mathrm{\Lambda}}_{0}^2)/4 = \tilde{G}_{\text{S}}^2 + F_{\text{S}}^2}$ and ${\hat{F}_{\text{S}} = F_{\text{S}} \exp[ \hat{X}_{30} i \chi ]}$, where ${\tilde{G}_{\text{S}} = r_y g(\omega) + \omega L^2 / D r_y}$ and ${F_{\text{S}} = r_y \mathcal{T}_{y}^2 f(\omega)}$. Equation~(\ref{g_0}) for~$\hat{g}_{0}$ describes the singlet component induced in the n~wire due to proximity effect. In the limit of a high transparency of the S/n interface (${r_{y} \to \infty}$), the matrix~$\hat{g}_{0}$ is close to the Green's function of the BCS superconductor. If the transparency is small (${r_{y} \ll 1}$), the matrix~$\hat{g}_{0}$ describes a singlet condensate with an amplitude which is not small at small energies ${\epsilon = i \omega \sim \epsilon_{\text{sub}}}$, where~$\epsilon_{\text{sub}}$ is the magnitude of a subgap induced by proximity effect. It is determined from the equation ${\mathcal{E}(-i \epsilon_{\text{sub}}) = 0}$ and equals ${\epsilon_{\text{sub}} = r_{y} E_{\text{Th}}}$, where ${E_{\text{Th}} = D/L^{2}}$ is the Thouless energy.

For the first correction we obtain, again using the normalization condition~Eq.~(\ref{eq:norm_cond}),
\begin{equation}
    \hat{g}_{1} = \frac{1}{2 \mathcal{E}} \big( \delta \hat{\mathrm{\Lambda}} - \hat{g}_{0} \cdot \delta \hat{\mathrm{\Lambda}} \cdot \hat{g}_{0} \big) \,. \label{6}
\end{equation}
It determines the Green's function in the structures shown in Figs.~\ref{fig:setup}~(a) and~\ref{fig:setup}~(b), but in order to understand the nature of interaction between the triplet and singlet components, we first consider the simpler system shown in Fig.~\ref{fig:setup}~(b).

\section{TS contact.}
\label{sec:TS_contact}

For simplicity, we assume that ${\chi_{\text{L}} = 0}$ and ${\alpha_{\text{L}} = 0}$. This choice is justified since the right S$_{\text{m}}$~superconductor is disconnected from the n~wire [see Fig.~\ref{fig:setup}~(b)].

In this case, the matrix~$\delta \hat{\mathrm{\Lambda}}$ has the form
\begin{equation}
\delta \hat{\mathrm{\Lambda}} = 2 r_x \mathcal{U}_{\text{L}} \big[ g_{+} ( \hat{X}_{30} + \zeta \hat{X}_{03} ) + f_{-}  \hat{X}_{m}(\zeta) \big] \,, \label{eq:Lambda_RL}
\end{equation}
where~$\hat{X}_{m}(\zeta)$ is given by
\begin{equation}
\hat{X}_{m}(\zeta) = \begin{cases}
                        \hat{X}_{x}(\zeta) \equiv \hat{X}_{11} - \zeta \hat{X}_{22} \,, & \text{if} \quad \mathbf{h} \parallel \hat{\mathbf{e}}_x \,, \\
                        \hat{X}_{y}(\zeta) \equiv \hat{X}_{12} + \zeta \hat{X}_{21} \,, & \text{if} \quad \mathbf{h} \parallel \hat{\mathbf{e}}_y \,,
                        \end{cases}
\end{equation}
i.e., depending on the way the coordinate system is chosen---either is the $x$~axis chosen as showing in the direction of~$\mathbf{h}$ or it is the $y$~axis---it defines a \emph{chirality} or \emph{handedness} ($m=x,y$) via the vector product ${\mathbf{h} \times \mathbf{n}_{\text{f}}}$ with~$\mathbf{n}_{\text{f}}$ determining the $z$~axis. In addition, the form of these matrices depends on the direction of the spin filters (${\zeta = \pm 1}$). In the case of the $x$\nobreakdash-chirality, the filter lets to pass Cooper pairs with spin up if ${\zeta = +1}$ and the S$_{\text{m}}$/n interface is transparent only for the Cooper pairs with spin down if ${\zeta = -1}$, and vice versa for the $y$\nobreakdash-chirality. One can show that terms given by~$\hat{X}_{x}(+1)$ describe correlations of the form ${\propto \langle \hat{c}_{\uparrow} \hat{c}_{\uparrow}(t)\rangle}$ while those given by~$\hat{X}_{x}(-1)$---of the form ${\propto \langle \hat{c}_{\downarrow} \hat{c}_{\downarrow}(t)\rangle}$; correspondingly describe terms given by~$\hat{X}_{y}(-1)$ correlations of the form ${\propto \langle \hat{c}_{\uparrow} \hat{c}_{\uparrow}(t)\rangle}$ while those given by~$\hat{X}_{y}(+1)$---of the form ${\propto \langle \hat{c}_{\downarrow} \hat{c}_{\downarrow}(t)\rangle}$.

Using the expression in Eq.~(\ref{eq:Lambda_RL}) for~$\delta \hat{\Lambda}$ we calculate the correction~$\hat{g}_1$ from Eq.~(\ref{6}),
\begin{equation}
\hat{g}_{1} = \frac{1}{2 \mathcal{E}^{3}} ( \delta \hat{g}_{\text{DoS}} + \hat{g}_{\text{s}} + \hat{g}_{\text{tr}} + \hat{g}_{M} ) \,. \label{11a}
\end{equation}
Here, ${\delta \hat{g}_{\text{DoS}} = g_{+} F_{\text{S}}^{2} \hat{X}_{30}}$ describes a correction to the density of states of the n~wire due to proximity effect in the S/n~system; the second term, ${\hat{g}_{\text{s}} = -2 \tilde{G}_{\text{S}} F_{\text{S}} \cos \chi \hat{X}_{10}}$ corresponds to the singlet component induced in the n~wire.

The most important term~$\hat{g}_{\text{tr}}$ describing the triplet Cooper pairs with a nonzero projection of the total spin onto the $z$\nobreakdash-axis equals
\begin{equation}
\hat{g}_{\text{tr}} = f_{-} \big[ (2 \tilde{G}_{\text{S}}^{2} + F_{\text{S}}^{2} ) \hat{X}_{m}(\zeta) - F_{\text{S}}^{2} \exp (2i \chi \hat{X}_{30}) \cdot \hat{X}_{m}(-\zeta) \big] \,. \label{11b}
\end{equation}
It describes the triplet components of different processes. The first term in the square brackets is the ``incident'' triplet component. The second term arises due to the interaction of the ``incident'' triplet component [$\propto f_{-} \hat{\mathrm{X}}_{m}(\zeta)$] with the singlet one induced in the n~wire from the superconductor~S. For the Cooper pairs with the chirality, say, ${m = x}$, this term can be written as
\begin{equation}
f_{-} F_{\text{S}}^{2} \big[ \cos (2 \chi) \hat{X}_{x}(-\zeta) + \zeta \sin (2 \chi) \hat{X}_{y}(-\zeta) \big] \,.  \label{11c}
\end{equation}
As discussed above, the matrices~$\hat{X}_{x}(\pm 1)$ and~$\hat{X}_{y}(\mp 1)$ describe the spin-up (respectively, spin-down) triplet Cooper pairs with different chirality.

The obtained results can be understood as a ``scattering'' of the triplet Cooper pairs by the singlet ones which leads to creation of new triplet Cooper pairs with another chirality and spin direction. We emphasize that this is the main result of this Paper, namely the fact that presence of the singlet condensate leads, i.a., to appearance of triplet Cooper pairs with opposite spin relative to the injected ones. We show in Section~\ref{sec:TST_Contact} that this fact has concrete physical consequences leading, e.g., to occurrence of current in a junction, where the current vanishes in case of missing singlet condensate.

The last term in Eq.~(\ref{11a}) describes the magnetic moment~$\mathbf{M}$ induced in the n~wire,
\begin{equation}
\hat{g}_{M}(\zeta )= -2 f_{-} \tilde{G}_{\text{S}} F_{\text{S}} \hat{X}_{30} \cdot
\begin{cases}
\hat{X}_{01} \cos \chi + \zeta \hat{X}_{02} \sin \chi \,, & \text{$x$\nobreakdash-chirality} \,, \\
\hat{X}_{02} \cos \chi - \zeta \hat{X}_{01} \sin \chi \,, & \text{$y$\nobreakdash-chirality} \,,
\end{cases}
\end{equation}

In case of the $x$\nobreakdash-chirality, the first term ($\propto \cos \chi $) determines the magnetic moment aligned parallel to the $x$~axis, while the second term ($\propto \sin \chi$) describes the magnetic moment along the $y$~axis. Note that the magnetic moment~$\mathbf{M}$ arises due to polarization of the singlet Cooper pairs by the triplet component. The triplet Cooper pairs are oriented in the $z$~direction, but the induced magnetic moment lies in the $(x,y)$~plane.

A similar effect arises in an S/F~bilayer with an uniform magnetization ${\mathbf{M}_{\text{F}} = (0, 0, M_{\text{F}})}$~\cite{Bergeret_Volkov_Efetov_2004,BVErmp,Bergeret_et_al_2005,Kharitonov_Volkov_Efetov_2006}. In the latter system, the triplet component with zero projection of the total spin~$\mathbf{S}$ on the $z$~axis, i.e., ${\mathbf{S} = (S_{x}, S_{y}, 0)}$, is induced in the F~film due to the proximity effect, and penetrates into the S~film due to the inverse proximity effect. This component polarizes the singlet Cooper pairs in the superconductor~S so that the magnetic moment~$\mathbf{M}$ arises in the S~film in the direction perpendicular to the $(x,y)$~plane, i.e., in the $z$~direction. This leads to screening (or anti-screening) of the magnetic moment of the ferromagnet~F which has been observed experimentally~\cite{Salikhov_et_al_2009,Xia_et_al_2009,Asulin_et_al_2009} (see also the recent paper Ref.~\onlinecite{Kalcheim_et_al_arXiv_2015}). Note that the magnetic moment caused by the LRTC has been calculated in other systems, where no filtering takes place, e.g., in Refs.~\onlinecite{Pugach_Buzdin_2012,Hikino_Yunoki_2015}.

As regards the currents in the considered TS~contact, inserting the solution given by Eq.~(\ref{11a}) into Eqs.~(\ref{2}), (\ref{4}),~and~(\ref{4a}), we obtain the rather anticipated results
\begin{align}
I_{\text{Q}} &= 0 \,, \label{eq:I_Q_TS} \\
I_{\text{sp}} &= 0 \,, \label{eq:I_sp_TS}
\end{align}
i.e., both currents vanish and there is no transfer of singlet Cooper pairs from the singlet superconductor into the ``magnetic'' superconductors, neither of triplet Cooper pairs vice versa.

\section{TST contact.}
\label{sec:TST_Contact}

As a consequence of ``scattering'' of the triplet Cooper pairs by the singlet condensate leading to creation of new triplet Cooper pairs with another chirality and spin direction, Eqs.~(\ref{11a})\nobreakdash--(\ref{11c}), it may be expected that there appears a finite current in a system, where there is none if the singlet condensate is absent. To be specific, we consider a structure consisting of two ``magnetic'' superconductors~S$_{\text{m}}$ connected by a normal wire which is attached to the superconductors via the antiparallel oriented filters. This structure has been considered in Ref.~\onlinecite{Moor_2015_arxiv} and it has been found there that, in this case, spin and charge Josephson currents vanish. Attaching an additional singlet superconductor to the normal wire we obtain the TST contact with antiparallel oriented filters as sketched in Fig.~\ref{fig:setup}~(a), thus introducing singlet condensate into the junction.

\begin{figure}
  \centering
  \includegraphics[width=1.0\columnwidth]{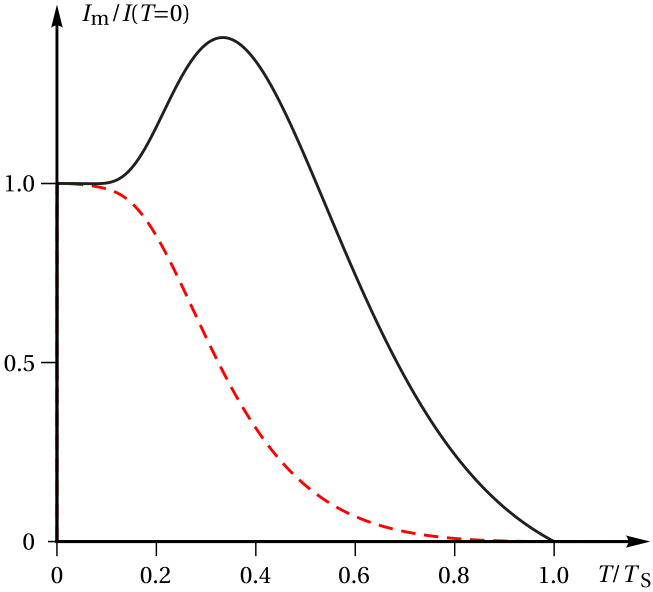}
  \caption{The coefficient~$I_{\text{m}}$ in the Josephson current [Eqs.~(\ref{eq:I_Q_3_terminal}) and~(\ref{eq:I_sp_3_terminal})] on temperature for different values of the exchange field~$h$, i.e., $h = 0.5 \Delta_{\text{T}}$ (red dashed curve) respectively $h = 2.0 \Delta_{\text{T}}$ (black solid curve), where~${\Delta_{\text{T}}}$ is the superconducting order parameter in the ``magnetic'' superconductor at ${T = 0}$. The current vanishes at ${T = T_{\text{S}}}$, where~$T_{\text{S}}$ is the superconducting transition temperature in the singlet superconductor assumed to be lesser than that of the ``magnetic'' superconductors. Note that in the case of $h = 2.0 \Delta_{\text{T}}$, the critical coefficient~$I_{\text{m}}$ is a non-monotonic function of the temperature and one observes an enhancement of~$I_{\text{m}}$ in a range of temperatures, resembling the results obtained in Refs.~\onlinecite{Eschrig08,Eschrig_2009}.} \label{fig:Current_on_T}
\end{figure}

From Eqs.~(\ref{2}), (\ref{4}),~and~(\ref{4a}), with the solution given by Eq.~(\ref{6}), we obtain, at the interface of the right triplet superconductor non vanishing Josephson and spin currents with a rather unusual phase dependence,
\begin{align}
I_{\text{Q}} &= I_{\text{m}} \sin (\chi_{\text{L}} + \chi_{\text{R}} - 2 \chi + \zeta_{-} \alpha) \,, \label{eq:I_Q_3_terminal} \\
I_{\text{sp}} &= \zeta_{-} \mu_{\text{B}} e^{-1} I_{\text{m}} \sin (\chi_{\text{L}} + \chi_{\text{R}} - 2 \chi + \zeta_{-} \alpha) \,, \label{eq:I_sp_3_terminal}
\end{align}
with ${\zeta_{-} = (\zeta_{\text{R}} - \zeta_{\text{L}})/2 = \pm 1}$ corresponding to the case when the right filter passes spin-up (respectively, spin-down) Cooper pairs.  We remind that the filters are assumed to be oriented antiparallel, ${\zeta_{\text{L}} = - \zeta_{\text{R}}}$ (i.e., int this case ${\zeta_{-} = \zeta_{\mathrm{R}}}$), and, that the case of ${\alpha = 0}$ corresponds to equal chiralities whereas ${\alpha = \pi/4}$---to the case of opposite chiralities of Cooper pairs participating in transport.
In Eqs.~(\ref{eq:I_Q_3_terminal}) and~(\ref{eq:I_sp_3_terminal}), the critical current~$I_{\text{m}}$ equals
\begin{equation}
I_{\text{m}} = I_{\text{c}} r_{\text{R}} r_{\text{L}} \sum f_{-}^{2} F_{\text{S}}^{2} \mathcal{E}^{-3} \big/ \sum f^{2} \mathcal{E}^{-2} \,,
\end{equation}
where the sum is taken over the Matsubara frequencies and~$I_{\text{c}}/r_{\text{S}}^2$ is the critical current of a standard Josephson S/n/S junction with the transparency~$r_{\text{S}}$. If the S$_{\text{m}}$/n/S$_{\text{m}}$~junction is not connected with the superconductor~S, the critical current~$I_{\text{m}}$ is zero (one has to set~${F_{\text{S}} = 0}$). This fact is clearly demonstrated in Fig.~\ref{fig:Current_on_T}, where we plot the temperature dependence of the critical current~$I_{\text{m}}(T)$. The current~$I_{\text{m}}(T)$ turns to zero when the temperature~$T$ exceeds the critical temperature of the singlet superconductor. Moreover, in Fig.~\ref{fig:Phase_Dependence} we present a sketch of the current--phase relation as obtained in Eqs.~(\ref{eq:I_Q_3_terminal}) and~(\ref{eq:I_sp_3_terminal}) for a fixed value of~$\alpha$ and~$\phi$ representing the total phase difference, i.e., for ${\phi = \chi_{\text{L}} + \chi_{\text{R}} - 2 \chi}$.

\begin{figure}
  \centering
  \includegraphics[width=0.8\columnwidth]{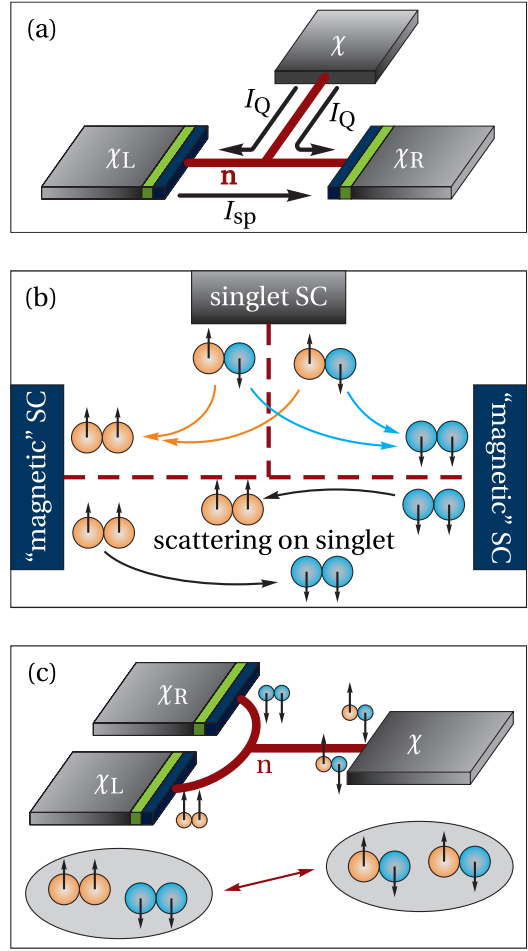}
  \caption{(a)~The Josephson charge and spin currents in a TST contact [cf.~Fig.~\ref{fig:setup}~(a)] as calculated in Eqs.~(\ref{eq:I_Q_3_terminal}) and~(\ref{eq:I_sp_3_terminal}) with subsequent discussion (see main text). The charge current~$I_{\text{Q}}$ flows from the singlet to the ``magnetic'' superconductors with spin filters. There is no spin transport between those terminals since the spin current~$I_{\text{sp}}$ flows between the triplet terminals only. The spin-polarized condensates move in the n~wires each towards the corresponding terminals in opposite direction, ${I_{\text{Q}}^{\text{R}} = - I_{\text{Q}}^{\text{L}}}$. Since spin-up spin current moving to the left corresponds to spin-down spin current moving to the right (and vice versa), the spin current at the right interface has the same direction as that at the left interface, ${I_{\text{sp}}^{\text{R}} = I_{\text{sp}}^{\text{L}}}$. (b)~Detailed resolution in terms of triplet and singlet Cooper pairs participating in charge and spin transport. The interaction between the singlet and triplet condensates leads to dissolution of two singlet Cooper pairs from the singlet superconductor into two triplet Cooper pairs moving towards the corresponding filter interface which lets to pass spin-up or spin-down electrons only. Also, the triplet Cooper pairs stemming from the corresponding ``magnetic'' superconductor are transformed upon interaction into triplet Cooper pairs with opposite spin direction [see Eq.~(\ref{11b}) and discussion in Section~\ref{sec:TS_contact}]. Here, the spin-up triplet Cooper pairs move to the left interface, whereas the spin-down triplet Cooper pairs---to the right with same velocity, thus yielding a net spin current but zero charge current. (c)~The expression Eq.~(\ref{eq:I_Q_3_terminal}) for the Josephson current can be understood as a Josephson relation for the case of two superconductors, one of which is a usual singlet one, and the second represents a combination of two triplet superconductors. In this case, the Josephson effect is essentially a four-fermion process.} \label{fig:Jos_Spin_Currents}
\end{figure}

Equations~(\ref{eq:I_Q_3_terminal}) and~(\ref{eq:I_sp_3_terminal}) represent the main result of the present Section. They determine the Josephson and spin currents in the system under consideration. The sign of the critical current is negative ($\pi$\nobreakdash-junction) because the function ${f_{-} = [f(\omega+ih) - f(\omega-ih)]/2}$ is purely imaginary. The dependence of the currents on the phases of the superconductors~S and~S$_{\text{m}}$ is purely sinusoidal, but this dependence is rather unusual---there is no dependence purely on the phase differences ${\chi_{\text{L}} - \chi}$ nor ${\chi_{\text{R}} - \chi}$. The current-phase relation, Eqs.~(\ref{eq:I_Q_3_terminal}) and~(\ref{eq:I_sp_3_terminal}), is similar to the corresponding dependence for an S$_{\text{M}}$/S/S$_{\text{M}}$ system, where superconductors~S$_{\text{M}}$ are coupled via Majorana fermions.\cite{Liang_et_al_2011} There is however a significant difference---the current~$I_{\text{Q}}$ in Ref.~\onlinecite{Liang_et_al_2011} is proportional to~${I_{\text{Q}} \simeq \sin [(\chi_{\text{L}} + \chi_{\text{R}})/2 - \chi]}$ unlike the dependence in Eq.~(\ref{eq:I_Q_3_terminal}). Thus, in the case considered by Jiang~\emph{et al.}, the current is transfered by single singlet Cooper pair which is split into two Majorana fermions. In our case, two singlet Cooper pairs passing through the n/S interface are divided into two triplet Cooper pairs with antiparallel total spins propagating in opposite directions (to the right and left~S$_{\text{m}}$ superconductors), cf.~Fig~\ref{fig:Jos_Spin_Currents}~(b)~and~(c), resulting in a different prefactor in front of the total phase difference.

At the interface of the left triplet superconductor, the Josephson charge current has opposite sign, whereas the spin current has the same sign as compared to those at the right interface, i.e., ${I_{\text{Q}} = I_{\text{Q}}^{\text{R}} = - I_{\text{Q}}^{\text{L}}}$, whereas ${I_{\text{sp}} = I_{\text{sp}}^{\text{R}} = I_{\text{sp}}^{\text{L}}}$. As sketched in Fig.~\ref{fig:Jos_Spin_Currents}~(a), by means of the current conservation it follows then that a charge current twice the calculated~$I_{\text{Q}}$ flows through the interface of the singlet superconductor. Correspondingly, there is no net spin current into or out the singlet superconductor. Note that the nature of the charge currents at different interfaces is essentially distinct. While at the interface of the singlet superconductors there exists charge current of singlet Cooper pairs, at the interfaces of the triplet superconductors, one observes currents of triplet Cooper pairs with correspondingly oriented spin projection, as follows from considerations on scattering processes in the TS~contact. Equivalently, one can view this as charge current carried by singlet Cooper pairs through the S/n interface being separated in the n~wire into two currents which are opposite to each other and are carried by triplet Cooper pairs with opposite spin direction to left, respectively, right terminal with corresponding filter.

As contrasted to results of Ref.~\onlinecite{Moor_2015_arxiv}, where we found that, in the case ${\alpha = 0}$, the currents vanish in the junction consisting of triplet superconductors with filters passing oppositely oriented Cooper pairs, now, they ``talk'' to each other via the singlet superconductor. It is important that both---the charge and the spin currents---are finite only if all three terminals are attached to the normal wire, ${r_{\text{R},\text{L}} \neq 0}$ and ${F_{\text{S}} \neq 0}$.

The interaction of the singlet and triplet Cooper pairs opens the way for the spin current to flow between the triplet terminals and for the charge current between the singlet terminal and the triplet ones.

Note that the both currents are proportional to each other, ${I_{\text{sp}} = \zeta_{-} \mu_{\text{B}} e^{-1} I_{\text{Q}}}$, which is interesting from the point of view of applications of the considered setup in superconducting spintronics,~\cite{Linder_Robinson_2015,Eschrig_Reports_2015} since it allows for controlling and monitoring the spin current directly via the Josephson current in an experimentally accessible way.

\section{Experiment proposal.}

\begin{figure}
  \centering
  \includegraphics[width=1.0\columnwidth]{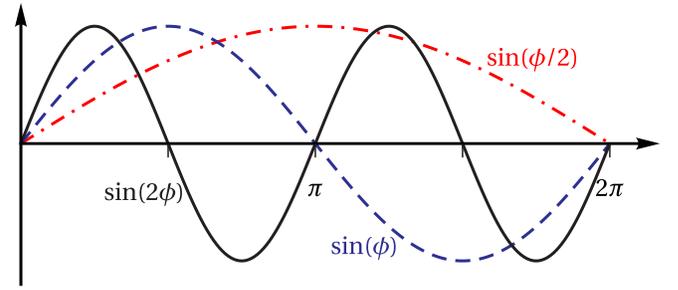}
  \caption{The current-phase relation of the Josephson spin and charge currents obtained in Eqs.~(\ref{eq:I_Q_3_terminal}) and~(\ref{eq:I_sp_3_terminal}) is a $\pi$-periodic function (solid black curve) as contrasted, e.g., the $2\pi$-periodicity in a usual Josephson contact (blue dashed curve), or the $4\pi$-periodic Josephson effect of a Majorana-type condensate (red dash-dotted curve).} \label{fig:Phase_Dependence}
\end{figure}

Perhaps, the simplest way to observe the predicted effect is to use the superconductor~S with a critical temperature~$T_{\text{S}}$ lesser than the critical temperatures of the superconductors~S$_{\text{m}}$, ${T_{\text{mR}} = T_{\text{mL}} \equiv T_{\text{m}}}$. Then, the dc Josephson current~$I_{\text{J}}$ is finite at ${T < T_{\text{S}}}$ and turns to zero at temperatures higher than~$T_{\text{S}}$, see Fig.~\ref{fig:Current_on_T}. On the other hand, the current~$I_{\text{J}}$ is finite if the filter axes are parallel to each other at any temperature less than~$T_{\text{m}}$, in which case ${I_{\text{J}} \neq 0}$ even in absence of the singlet superconductor~S~\cite{Moor_2015_arxiv}.

A more sophisticated experiment would involve circuits of superconducting loops additionally connecting the terminals:
\begin{itemize}
\item If the left triplet superconductor~S$_{\text{m,L}}$ is connected with the singlet superconductor~S, the phases of both superconductors are equal (in the absence of a magnetic flux in the loop connecting the superconductors), ${\chi = \chi_{\text{L}}}$. In this case, the Josephson current flows from~S to S$_{\text{m,R}}$, and the current-phase relation has the usual form: ${I_{\text{Q}} = I_{\text{c}} \sin(\chi_{\text{R}} - \chi)}$. If there is a magnetic flux~$\Phi$ in the loop, then the difference ${\chi_{\text{R}} - \chi}$ should be replaced by ${\chi_{\text{R}} - \chi + 2 \pi \Phi/\Phi_{0}}$, where~$\Phi_{0}$ is the magnetic flux quantum;
\item If superconductors~S$_{\text{m,R}}$ and ~S$_{\text{m,L}}$ are connected, then ${\chi_{\text{R}} = \chi_{\text{L}}}$ (in absence of~$\Phi$) and we obtain for the current the relation ${I_{\text{Q}} = I_{\text{c}} \sin [2 (\chi_{\text{R}} - \chi)]}$, i.e., the period of the current oscillations as a function of the phase difference equals~$\pi$, see Fig.~\ref{fig:Phase_Dependence} with ${\phi = \chi_{\text{R}} - \chi = \chi_{\text{L}} - \chi}$;
\item Assume that the left triplet superconductor~S$_{\text{m,L}}$ is disconnected from the circuit. Then, no supercurrent can flow through the branch S/n/S$_{\text{m,R}}$ as well. This means that the phase~$\chi_{\text{L}}$ of the left triplet superconductor~S$_{\text{m,L}}$ is adjusted in such a way that ${\chi_{\text{R}} + \chi_{\text{L}} - \chi = 0}$.
\end{itemize}

\section{Conclusion.}

Using the quasiclassical approach we have shown that interaction of a polarized $s$\nobreakdash-wave triplet component and a singlet one results in creation of triplet Cooper pairs with opposite spin direction or of different chiralities. Such spin transformation leads to interesting effects in magnetic Josephson junctions.

Considering the dc Josephson effect in a multiterminal Josephson contact of the {S$_{\text{m}}$/n/S/n/S$_{\text{m}}^{\prime}$~type with ``magnetic'' superconductors~S$_{\text{m}}$, that generate fully polarized triplet components, and a singlet superconductor~S with a phase~$\chi$, we determined the Josephson and spin currents in the case when the filter axes of the triplet superconductors~S$_{\text{m}}$ are oriented antiparallel to each other.

As contrasted to the case when the singlet superconductor is absent~\cite{Moor_2015_arxiv}, the currents are finite and show an unusual phase dependence on the phases~$\chi_{\text{L/R}}$ of superconductors~S$_{\text{m}}$, ${I_{\text{J},\text{sp}} \propto I_{\text{c}} \sin ( \chi_{\text{R}} + \chi_{\text{L}} - 2 \chi )}$, i.e., they do not depend on the difference, ${\chi_{\text{R}} - \chi_{\text{L}}}$, between the phases of the right and left triplet superconductors~S$_{\text{m}}$.

Interestingly, this expression represents a Josephson effect taking place between two usual singlet superconductors with phases~${\chi_{\text{R}} + \chi_{\text{L}}}$ and~$2 \chi$, respectively. Qualitatively, this is true since there occurs Josephson current between the singlet superconductor and the two triplet terminals which can be viewed as one singlet superconductor, Fig.~\ref{fig:Jos_Spin_Currents}~(c), but quantitatively this describes a doubling of the relative phase difference. Thus, the current-phase relation is a $\pi$\nobreakdash-periodic function as contrasted, e.g., the $2\pi$\nobreakdash-periodicity in a usual Josephson contact, or the $4\pi$\nobreakdash-periodic Josephson effect of a Majorana-type condensate~\cite{Kitaev_2001,Lutchyn_Sau_DSarma_2010}.

The Josephson currents vanish in case when one arbitrary terminal is disconnected from the heterostructure.

We discussed possibilities of experimental observation of the effect.

\acknowledgments

We appreciate the financial support from the DFG via the Projekt~EF~11/8\nobreakdash-2; K.~B.~E.~gratefully acknowledges the financial support of the Ministry of Education and Science of the Russian Federation in the framework of Increase Competitiveness Program of  NUST~``MISiS'' (Nr.~K2-2014-015).


%

\end{document}